\documentstyle[preprint,aps]{revtex}
\begin{document}
%
%
\preprint{$
\begin{array}{l}
\mbox{UR-1637}\\[-3mm]
\mbox{UB-HET-01-03}\\[-3mm]
\mbox{June~2001} \\ [5mm]
\end{array}
$}
\title{Direct Measurement of the Top Quark Charge at Hadron Colliders
\\[0.2cm]}
\author{U.~Baur\footnote{e-mail: baur@ubhex.physics.buffalo.edu}}
\address{Department of Physics,
State University of New York, Buffalo, NY 14260, USA\\[1.mm]}
\author{M.~Buice\footnote{e-mail: mabuice@uchicago.edu}}
\address{Department of Physics,
University of Chicago, 5640 S. Ellis Ave., Chicago, IL 60637,
USA\\[1.mm]}
\author{Lynne H.~Orr\footnote{e-mail: orr@pas.rochester.edu}}
\address{Department of Physics and Astronomy,
University of Rochester, Rochester, NY 14627, 
USA\\[1.mm]}
\maketitle
\tightenlines
%
%
\begin{abstract}
\baselineskip15.pt  
We consider photon radiation in $\bar tt$ events at the upgraded
Fermilab Tevatron and the CERN Large Hadron Collider (LHC) as a tool to
measure the electric charge of the top quark. We analyze the
contributions of $\bar tt\gamma$ production and radiative top quark
decays to $p\,p\hskip-7pt\hbox{$^{^{(\!-\!)}}$} \to \gamma\ell^\pm\nu
\bar bbjj$, assuming that both $b$-quarks are tagged. With 20~fb$^{-1}$
at the Tevatron, the possibility that the ``top quark'' discovered in
Run~I is actually an exotic charge $-4/3$ quark can be ruled out at the
$\approx 95\%$ confidence level.
At the LHC, it will be possible to
determine the charge of the top quark with an accuracy of about 10\%.

\end{abstract}
\newpage
%
%
\section{Introduction}
It is widely believed that the new particle discovered at Fermilab in
1995~\cite{topcdf,topd0} is the long sought top quark. 
Its behavior is certainly consistent with Standard Model (SM)
expectations for top, but many of its 
properties are still only poorly known.  In
particular, the top quark's electric charge --- one of the most fundamental 
quantities characterizing a particle --- has not been measured yet.
It still remains not only to confirm that the discovered  quark has 
charge $2/3$ (and hence the expected SM quantum numbers), but also to
measure the strength of its electromagnetic (EM) coupling to rule out
anomalous contributions to its EM interactions.

Indeed, one alternative interpretation has not yet been ruled out:
that the new particle is a charge $-4/3$ quark. 
In the published top quark analyses of the
CDF and D\O\ Collaborations (see Ref.~\cite{topreview} for a review), 
the correlations of the $b$-quarks and the $W$
bosons in $\bar pp\to\bar tt\to W^+W^-\bar bb$ are not determined. As a
result, there is a twofold ambiguity in the pairing of $W$ bosons and
$b$-quarks, and, consequently, in the electric charge assignment of the
``top quark''. Besides the Standard Model (SM) assignment, $t\to W^+b$,
$''t''\to W^-b$ is also conceivable, in which case the ``top quark'' would
actually be an exotic quark with charge $q=-4/3$. 

Interpreting the particle found at Fermilab as a charge $-4/3$ quark 
is consistent with current
precision electroweak data. Current $Z\to\ell^+\ell^-$ and
$Z\to\bar bb$ data can be fit with a top quark of mass
$m_t\approx 270$~GeV, provided that the right-handed $b$-quark mixes
with the isospin $+1/2$ component of an exotic doublet of charge $-1/3$
and $-4/3$ quarks, $(Q_1,Q_4)_R$~\cite{ccm}. If the top quark would have
a mass of $m_t\approx 270$~GeV, it would have escaped detection in Run~I
of the Tevatron. In this scenario, the particle discovered in Run~I is
the $Q_4$.  A direct measurement of the top quark charge would 
provide a definitive test of this interpretation.

There are several techniques to determine the electric charge of the top
quark, $q_{top}$, in future collider experiments. 
At a linear $e^+e^-$ collider, the
full set of form factors describing the most general $\gamma\bar tt$
vertex function compatible with Lorentz invariance can be
probed in $e^+e^-\to\bar tt$~\cite{hioki} (similarly for a muon collider). 
A photon-photon collider would provide even more precision via the
two electromagnetic vertices in $\gamma\gamma\to\bar tt$.
The  status of future 
lepton and photon colliders is unfortunately still somewhat uncertain, but
in the meantime top quarks will be produced copiously at the 
Fermilab Tevatron and the CERN LHC.
At these hadron colliders, $\bar
tt$ production is so dominated by the QCD processes $\bar
qq\to g^*\to\bar tt$ and $gg\to\bar tt$ that  a measurement of the
$\gamma\bar tt$ form factors via $q\bar q\to\gamma^*\to\bar tt$ is hopeless. 
Instead, one can measure the top charge by measuring the charges of
its decay products, the final state $b$-jets and $W$ bosons.
One can further 
attempt to determine not only the top quark's charge but also the EM
coupling strength by directly studying  top's electromagnetic interactions
through photon radiation in $\bar tt$ events. 

The first method --- measuring the top charge by reconstructing the 
charges of its decay products --- is difficult to realize for $\bar tt\to$~all
jet decays; however for di-lepton decays, $\bar
tt\to\ell^+\nu\ell^-\bar\nu\bar bb$, and for semileptonic decays, $\bar
tt\to\ell^\pm\nu jj\bar bb$, it should be feasible. 
The $b$-jet charge can be determined from a measurement of
the charges associated with the tracks in the jet. 
A preliminary measurement of the $b$-jet charge using 
Run~I CDF data~\cite{velev} shows a slight preference for the SM top charge
assignment. The direct
measurement of the $b$-jet charge has the disadvantage that many
tagged $b$-quarks are needed to obtain a statistically
significant result. In a given event, missing or out-of-cone 
tracks can distort the measurement. Information on the $b$-jet 
charge can also be obtained from the charge-sign of the lepton, 
$\ell_b=e,\,\mu$, in events where the
$b$-quark is identified via a soft lepton tag, i.e. where the
$b$-quark decays semileptonically, $b\to c\ell_b\nu$. In the absence of 
$B_d-\overline{B_d}$ and $B_s-\overline{B_s}$ mixing, the charge-sign of
the lepton is directly correlated with the charge-sign of the parent
$b$-quark. The difficulties associated with using soft lepton tagged 
$b$-quarks are
low efficiency due to the small $b\to c\ell\nu$ branching ratio
and the presence of wrong sign leptons originating from 
$B_d-\overline{B_d}$ and
$B_s-\overline{B_s}$ mixing, which one has to correct for. In 
addition, if only the charge of the $b$
decay lepton is measured, only the sign but not the magnitude of the
charge of the $b$-quark is determined. 

In this paper, we explore the possibility of measuring the electric
charge of the top quark at hadron colliders through photon radiation 
in $\bar tt$ events, assuming that the EM coupling strength takes its SM
value\footnote{This is required by electromagnetic gauge invariance for
on-shell photons radiated in $\bar tt$ events.}. Because top quarks can radiate 
photons in both top quark production and top decay, we consider both 
processes:
\begin{equation}
p\,p\hskip-7pt\hbox{$^{^{(\!-\!)}}$} \to \bar tt\gamma\qquad{\rm and}\qquad
p\,p\hskip-7pt\hbox{$^{^{(\!-\!)}}$} \to \bar tt,~t\to Wb\gamma. 
\end{equation}
We concentrate on the lepton+jets mode\footnote{Although
di-lepton
events, $p\,p\hskip-7pt\hbox{$^{^{(\!-\!)}}$} \to
\gamma\ell^+\nu\ell^-\bar\nu\bar bb$, are cleaner than  
lepton+jets events,  their
branching fraction is about a factor~6 smaller than that of the
lepton+jets mode. The all-jets mode has a larger branching
ration than lepton+jets, but is plagued by a large QCD background
and, therefore, is also not considered here.},
\begin{equation}
p\,p\hskip-7pt\hbox{$^{^{(\!-\!)}}$} \to \gamma\ell^\pm\nu b\bar bjj,
\end{equation}
$\ell=e,\,\mu$, and assume that both $b$-quarks are tagged. 
We present results for the SM top quark and for comparison
we also consider a charge $-4/3$ quark of the type discussed
in~\cite{ccm}. 

Photon radiation in top quark events was not observed by the Tevatron
experiments in Run~I. In Run~II, an integrated
luminosity of $2-20~{\rm fb}^{-1}$ is envisioned, and a sufficient
number of $\gamma\ell^\pm\nu b\bar bjj$ events may be available to measure
the top quark charge. At the LHC, where the cross
section for $\bar tt$ production is more than a factor~100 larger than
at the Tevatron, one can hope for a determination of $q_{top}$ with a
precision of about 10\%.

Details of our calculation are given in Sec.~II. In Sec.~III we present
results of our numerical studies.  In addition to total cross sections,
distributions for
the transverse momentum  of the photon and the  separation
between the photon and the $b$-quarks in the pseudorapidity -- azimuthal
angle plane are found to be sensitive to $q_{top}$. In Sec.~IV we perform
a quantitative analysis of how well the charge of the top quark can be
measured at the Tevatron and the LHC. Finally, in Sec.~V, we present
our conclusions.

\section{Details of the Calculation}

Our calculation is carried out at the tree level. We assume that both
$b$-quarks are tagged with a combined efficiency of 40\%, which is
included in all our numerical results. Top quark and $W$
boson decays are treated in the narrow width approximation. In this
approximation, there are three contributions to
$p\,p\hskip-7pt\hbox{$^{^{(\!-\!)}}$} \to \gamma\ell^\pm\nu b\bar bjj$:
\begin{itemize}
\item Radiation in top production: 
$p\,p\hskip-7pt\hbox{$^{^{(\!-\!)}}$} \to \bar tt\gamma\to 
\gamma\ell^\pm\nu b\bar bjj$. 

\item Radiative top decay: $p\,p\hskip-7pt\hbox{$^{^{(\!-\!)}}$} \to 
\bar tt\to\gamma
W^+bW^-\bar b\to\gamma\ell^\pm\nu b\bar bjj$. Here either the $t$ or $\bar t$
quark decays radiatively. We denote these decays generically by $t\to
Wb\gamma$. 

\item Radiative $W$ decay: $p\,p\hskip-7pt\hbox{$^{^{(\!-\!)}}$} \to
\bar tt\to W^+bW^-\bar
b\to\gamma\ell^\pm\nu b\bar bjj$. In this case, one of the $W$ bosons
decays radiatively, $W\to\ell\nu\gamma$ or $W\to jj\gamma$. 
\end{itemize}

Both $\bar tt\gamma$ production and radiative top decays are sensitive
to $q_{top}$. On the other hand, radiative $W$ decays are not, and should be
suppressed by cuts. This can be achieved by requiring that
\begin{equation}
m(jj\gamma)>90~{\rm GeV\qquad and}\qquad m_T(\ell\gamma;p\llap/_T)>
90~{\rm GeV},
\label{eq:radw}
\end{equation}
where $p\llap/_T$ denotes the missing transverse momentum originating
from the neutrino in the event which is not observed, and $m(jj\gamma)$
is the invariant mass of the $jj\gamma$ system. 
$m_T(\ell\gamma;p\llap/_T)$ is the $\ell\gamma p\llap/_T$ cluster
transverse mass, which is given by
\begin{equation}
m_T^2(\ell\gamma;p\llap/_T)=\left(\sqrt{p_T^2(\ell\gamma)+m^2(\ell\gamma)}
+ p\llap/_T\right)^2-\left(\mathbf{p}_T(\ell\gamma)+
\mathbf{p\llap/}_T\right)^2\, ,
\end{equation}
where $p_T(\ell\gamma)$ and $m(\ell\gamma)$ are the transverse momentum
and the invariant mass of the $\ell\gamma$ system, respectively. For
$W\to\ell\nu\gamma$, the cluster transverse mass sharply peaks at the
$W$ boson mass, $M_W$. In the narrow width approximation, the cuts
listed in Eq.~(\ref{eq:radw}) completely eliminate the contributions from
radiative $W$ decays to the $\gamma\ell^\pm\nu b\bar bjj$ final
state. We thus ignore radiative $W$ decays in the following. 

The matrix elements for 
\begin{equation}
gg,~\bar qq\to \bar tt\gamma\to\gamma\ell^\pm\nu b\bar bjj
\end{equation}
are calculated using {\tt MADGRAPH}~\cite{tim} and the {\tt HELAS}
library~\cite{helas}, including the spin correlations for the subsequent
top decays. To compute the matrix elements for 
\begin{equation}
gg,~\bar qq\to \bar tt, \qquad t\to W^+ b\gamma, \qquad \bar t\to
W^- \bar b
\end{equation}
and its charge conjugate sibling, we use the crossed form of the $W\to
tb\gamma$ matrix elements of
Ref.~\cite{ds}. We ignore spin correlations in the decay of the $W$
boson in $t\to W^+b\gamma$. For the $\bar t\to W^-\bar b$ decay,
spin correlations are taken into account. All numerical results presented 
below are obtained summing over
electron and muon final states.

The top quark charge, $q_{top}$,
is treated as a free parameter in our calculation. The electric charge of the
$b$-quark, $q_b$, is related to $q_{top}$ by 
\begin{equation}
q_b= q_{top}-q_W, 
\end{equation}
where $q_W=\pm 1$ is the charge of the $W$ boson. The sign of $q_W$ depends on
whether the third component of the weak isospin of the ``top quark'' is
$T_3=+1/2$, as in the SM, or $T_3=-1/2$ as in the scenario of
Ref.~\cite{ccm}. We assume the $Wtb$ coupling to be of $V-A$ form and 
BR$(t\to Wb)=1$. The fermions originating from $W$ decays are assumed to be
massless. 

We note for future reference that the differential cross 
sections for $gg\to\bar tt\gamma$ and for $q\bar q,\,gg\to\bar
tt$ with one top quark decaying via $t\to Wb\gamma$ are invariant under
the transformation
\begin{equation}
q_{top}\to -q_{top}, \qquad q_W\to -q_W, \qquad q_b\to -q_b.
\label{eq:trans}
\end{equation}
Since interference effects between the incoming light quarks and the final
state top quarks in $q\bar q\to \bar tt\gamma$ are very small (they
contribute less than 1\% to the cross section), the cross
section for $p\,p\hskip-7pt\hbox{$^{^{(\!-\!)}}$} \to \gamma\ell^\pm\nu
b\bar bjj$ is to a very good approximation invariant under the
transformation~(\ref{eq:trans}). This will be relevant when we estimate
how well $q_{top}$ can be measured in future Tevatron and LHC experiments
(see Sec.~IV). 

For all our numerical simulations we have chosen the set of SM input
parameters to be: $m_t=175$~GeV~\cite{mtopcdf,mtopd0}, $m_b=5$~GeV,
$M_W=80.3$~GeV, $\sin^2\theta_W=0.23$
and $\alpha(M_W)=1/128$~\cite{lep}. We employ MRSR2 parton distribution
functions~\cite{mrs}, selecting the value of the factorization scale to
be $\mu_f=\sqrt{\hat s}$, where $\hat s$ is the squared parton center of
mass energy. Calculations are carried out for $p\bar p$ collisions at
$\sqrt{s}=2$~TeV (Tevatron) and $pp$ collisions at $\sqrt{s}=14$~TeV
(LHC). 

To simulate detector response, we impose the following common transverse
momentum, rapidity and separation cuts at the Tevatron and LHC:
\begin{eqnarray}
p_T(b) >15~{\rm GeV,} & \qquad & |\eta(b)|<2, \\\label{eq:top31}
p_T(\ell)>20~{\rm GeV,} & \qquad & |\eta(\ell)|<2.5,\\\label{eq:top32}
p_T(j_{1,2})>20~{\rm GeV,} & \qquad & |\eta(j_{1,2})|<2.5,\\\label{eq:top33}
p\llap/_T>20~{\rm GeV,} & \qquad & \Delta R(i,j)>0.4~{\rm for}~i\neq j, 
\label{eq:top35}
\end{eqnarray}
where $i,j=\gamma,\,\ell,\,b,\,\bar b,\,j_1,\,j_2$. $j_1$ and
$j_2$ denote the two jets originating from the hadronically decaying $W$
in the event. 
As mentioned above, we assume that both $b$ quarks are tagged 
(40\% combined efficiency)
so that $b$ and $\bar b$ are distinguishable from $j_1$ and $j_2$.
In addition, we require that
\begin{eqnarray}\label{eq:top4}
p_T(\gamma)>10~{\rm GeV,} & \qquad |\eta(\gamma)|<2.5 & \qquad {\rm 
at~the~Tevatron~and}\\
p_T(\gamma)>30~{\rm GeV,} & \qquad |\eta(\gamma)|<2.5 & \qquad {\rm 
at~the~LHC.} \label{eq:top5}
\end{eqnarray}
Here, $\eta$ is the pseudo-rapidity, and 
\begin{equation}
\Delta R(i,j)=\left [(\Delta\Phi(i,j))^2+(\Delta\eta(i,j))^2\right
]^{1/2}
\end{equation}
is the separation between two particles $i$ and $j$ in the 
pseudorapidity -- azimuthal angle plane. 

The transverse momentum cuts listed in Eqs.~(\ref{eq:top31})
--~(\ref{eq:top35}) and in Eq.~(\ref{eq:top5}) for the LHC are
sufficient for running at low luminosity, ${\cal L}=10^{33}\,{\rm
cm}^{-2}\,{\rm s}^{-1}$. For operation at the design luminosity of
${\cal L}=10^{34}\,{\rm cm}^{-2}\,{\rm s}^{-1}$, the $p_T$ cuts must be 
increased due to pile-up effects caused by the large number of
interactions per beam crossing. In the following we only consider 
$\gamma\ell^\pm\nu b\bar bjj$ production at the LHC in the low
luminosity environment. 

We do not consider any potential background processes here. The dominant 
background is expected to originate 
from $W\gamma + $~jets production and should be manageable in 
much the same way as the $W+$~jets background for $t\bar t$ production.

\section{Phenomenological Results}

In this section we present our results for total cross sections and various
photon distributions.  We look at contributions from individual subprocesses 
(radiative production, radiative decays) and compare results 
for the SM top quark to those
for an exotic  charge $-4/3$ top (as in Ref.~\cite{ccm}) in order to 
determine which quantities are sensitive to the value of $q_{top}$.

The events passing the cuts listed in Eqs.~(\ref{eq:radw})
and~(\ref{eq:top31}) --~(\ref{eq:top5}) can be split into three separate
samples, each designed to enhance one of the three subprocesses:  
(i) radiation
in top production (``$\bar tt\gamma$ selection cuts''), (ii) radiative top
decay with leptonically decaying $W$ (``$t\to Wb\gamma\to\ell\nu b\gamma$ 
selection cuts''), and (iii) radiative top decay with hadronically 
decaying $W$ (``$t\to Wb\gamma\to jj b\gamma$ selection cuts'').

The event sample for radiation in top production ($\bar tt\gamma$ selection 
cuts) is obtained by suppressing radiative top decay events.
We do this by selecting events which satisfy 
\begin{eqnarray}
m(b_{1,2}jj\gamma)>190~{\rm GeV} & \qquad {\rm and} & \qquad
m_T(b_{2,1}\ell^\pm\gamma;p\llap/_T) >190~{\rm GeV}.
\label{eq:top51}
\end{eqnarray}
Here, $b_1,\,b_2=b,\,\bar b$ and
$b_1\neq b_2$.  We note that by imposing the cuts on either $b$-quark 
combination,
we conservatively assume that the charges of the individual $b$ quarks have not
been determined. 

The total cross sections for $q_{top}=2/3$ and
$q_{top}=-4/3$, imposing the $\bar tt\gamma$ selection cuts of 
Eq.~(\ref{eq:top51}) and the cuts described in Sec.~II, are listed in
Table~\ref{tab:one}a for the Tevatron and the LHC. At the Tevatron, 
the $\bar
tt\gamma$ cross section is completely dominated by $q\bar q$
annihilation.  As a result, photon radiation off the initial state quarks
constitutes an irreducible background which limits the sensitivity of
the cross section to $q_{top}$. In contrast, at the LHC more than 75\%
of the $\bar tt\gamma$ cross section originates from gluon fusion.  Since
gluons do not radiate photons, the LHC cross section 
 scales approximately with $q^2_{top}$. Radiative top decays
contribute about 10 -- 20\% (1\%) to the cross section for $q_{top}=2/3$
($q_{top}=-4/3$) in this phase space region. At both the Tevatron
and LHC, increasing  $|q_{top}|$ increases the total cross section
with $\bar tt\gamma$ selection cuts.

\begin{figure}
\phantom{x}
\vskip 14.cm
\includegraphics{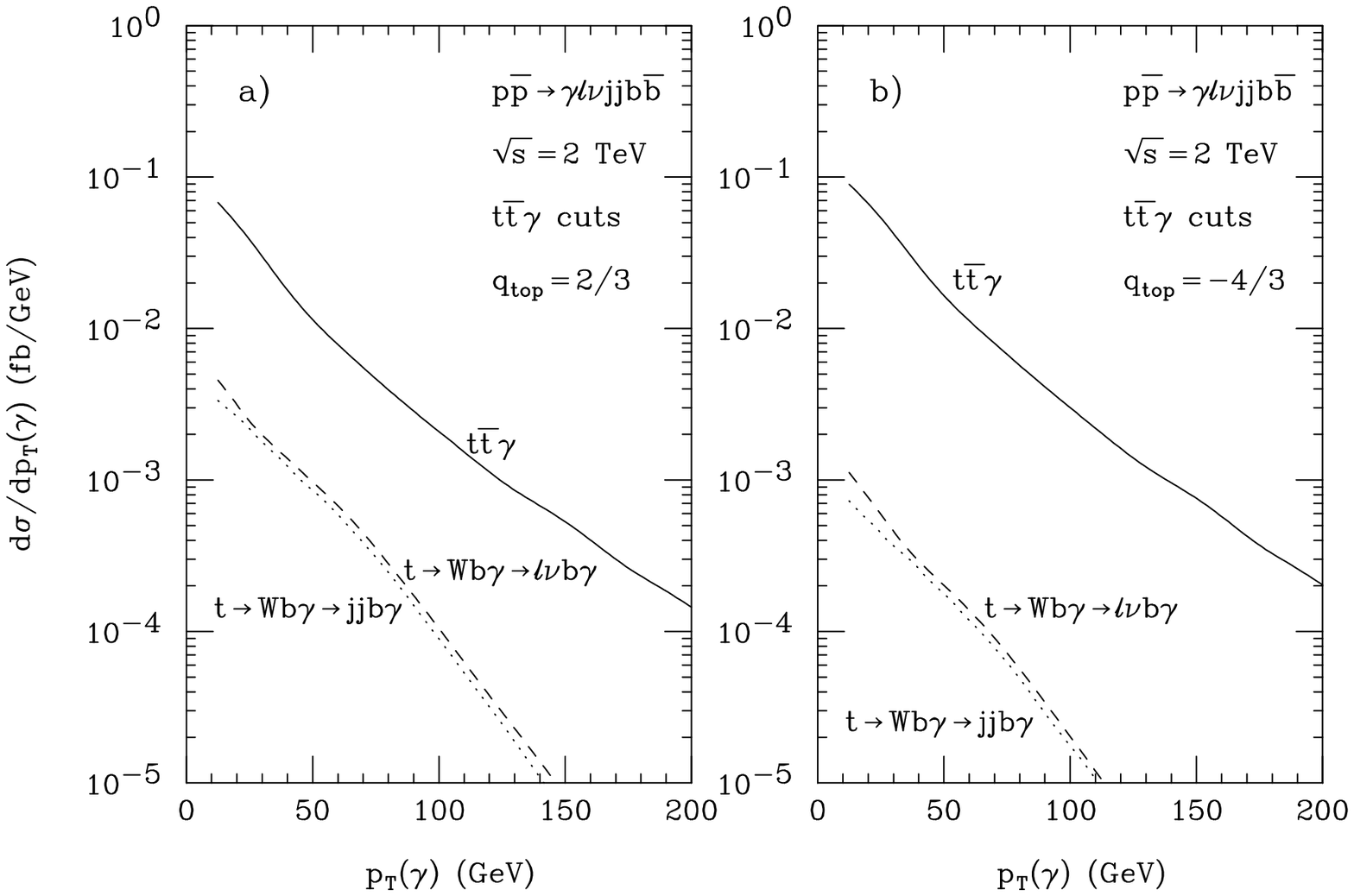}
\caption{The photon transverse momentum distribution for $p\bar p\to
\gamma\ell^\pm\nu b\bar bjj$ at the Tevatron, imposing $\bar tt\gamma$ 
selection cuts (see Eq.~(\ref{eq:top51})), for a) $q_{top}=2/3$ and b) 
$q_{top}=-4/3$. Shown are the $\bar tt\gamma$ (solid line), 
$\bar tt$, $t\to Wb\gamma\to\ell\nu b\gamma$ (dashed line), and $\bar
tt$, $t\to Wb\gamma\to jjb\gamma$ (dotted line) contributions. The 
additional cuts imposed are described in Sec.~II.}
\label{fig:fig0}
\end{figure}

\begin{figure}
\phantom{x}
\vskip 14.cm
\includegraphics{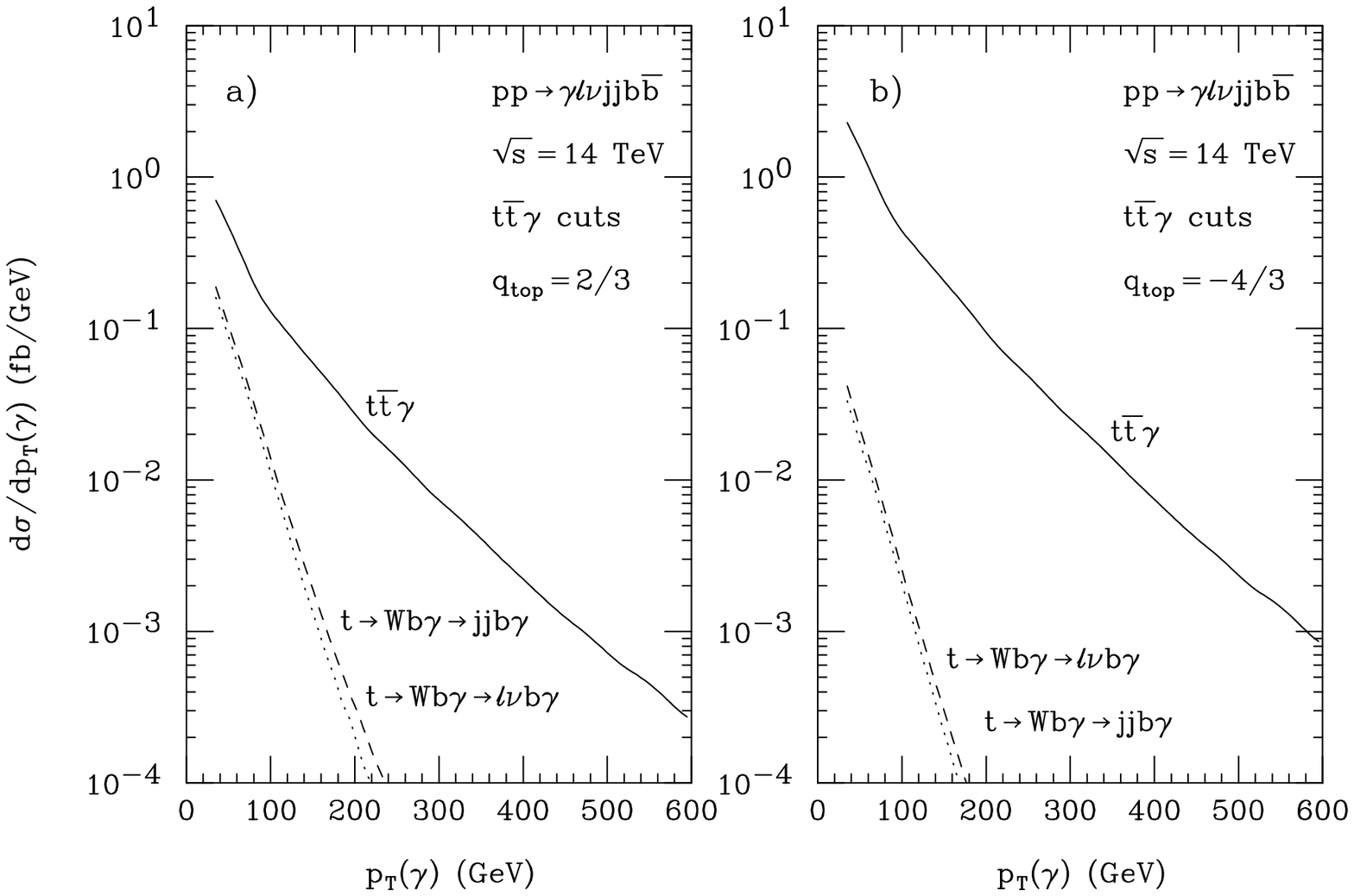}
\caption{The photon transverse momentum distribution for $pp\to
\gamma\ell^\pm\nu b\bar bjj$ at the LHC, imposing $\bar tt\gamma$ 
selection cuts (see Eq.~(\ref{eq:top51})), for a) $q_{top}=2/3$ and b) 
$q_{top}=-4/3$. Shown are the $\bar tt\gamma$ (solid line), 
$\bar tt$, $t\to Wb\gamma\to\ell\nu b\gamma$ (dashed line), and $\bar
tt$, $t\to Wb\gamma\to jjb\gamma$ (dotted line) contributions. The 
additional cuts imposed are described in Sec.~II.}
\label{fig:fig0a}
\end{figure}

The photon transverse momentum distributions for 
the individual contributions  to the 
$\gamma\ell^\pm\nu b\bar bjj$ differential cross section at the Tevatron
(LHC) for $\bar tt\gamma$ selection cuts are shown in Fig.~\ref{fig:fig0}
(Fig.~\ref{fig:fig0a}) for both
$q_{top}=2/3$ and $q_{top}=-4/3$, keeping the same vertical scale 
for both charges. The cross section from $\bar tt$
production followed by $t\to Wb\gamma\to\ell\nu b\gamma$ is 
about 20\% larger that from $p\bar p\to\bar tt$ followed by $t\to
Wb\gamma\to jjb\gamma$ over most of the $p_T$ range considered. At the 
LHC, the $p_T(\gamma)$ distribution for
the radiative top decay channels drops much faster than that for
$p\,p\hskip-7pt\hbox{$^{^{(\!-\!)}}$} \to \bar tt\gamma$.

A sample for radiative top
decay with leptonically decaying $W$ ($t\to Wb\gamma\to\ell\nu b\gamma$ 
selection cuts) is obtained by
requiring that
\begin{eqnarray}
m_T(b_{1,2}\ell^\pm\gamma;p\llap/_T) <190~{\rm GeV} & \qquad {\rm and} & 
\qquad m(b_{2,1}jj\gamma)>190~{\rm GeV} 
\label{eq:top6}
\end{eqnarray}
in addition to~(\ref{eq:top51}) being {\sl not} satisfied.  
These  cuts  enhance the contribution of the  
process $p\,p\hskip-7pt\hbox{$^{^{(\!-\!)}}$} \to t\bar t$, with 
$t\to Wb\gamma\to\ell\nu b\gamma$. 
Similarly, to get events with radiative top decay and hadronically decaying
$W$ ($t\to
Wb\gamma\to jjb\gamma$ selection cuts) we impose
\begin{eqnarray}
m_T(b_{1,2}\ell^\pm\gamma;p\llap/_T)
>190~{\rm GeV} & \qquad {\rm and} & \qquad 160~{\rm
GeV}<m(b_{2,1}jj\gamma)<190~{\rm GeV}
\label{eq:top7}
\end{eqnarray}
and require that~(\ref{eq:top51}) is not fulfilled.  
We thereby obtain an event sample in which 
the process  $p\,p\hskip-7pt\hbox{$^{^{(\!-\!)}}$} \to t\bar t$, with 
$t\to Wb\gamma\to jjb\gamma$ is enhanced. The cuts are imposed on either 
of the $b$-quark 
combinations, i.e. we assume that it has not been determined whether
the $b$ or the $\bar b$ originates from the radiative top quark decay.  
Since the finite experimental resolution of detectors significantly 
broadens the Breit-Wigner resonance for $t\to Wb\gamma\to jjb\gamma$, we
have chosen a relatively large window for $m(b_{2,1}jj\gamma)$ around
$m_t$. Requiring that events do not satisfy~(\ref{eq:top51})
significantly improves the efficiency of the $t\to Wb\gamma$ selection
cuts. 

The cross sections for $p\,p\hskip-7pt\hbox{$^{^{(\!-\!)}}$} \to
\gamma\ell^\pm\nu b\bar bjj$ imposing $t\to Wb\gamma\to\ell\nu b\gamma$
(Eq.~(\ref{eq:top6})) and $t\to Wb\gamma\to jjb\gamma$ selection cuts 
(Eq.~(\ref{eq:top7})) are shown in Table~\ref{tab:one}b
and~\ref{tab:one}c for $q_{top}=2/3$ and $q_{top}=-4/3$. The total
cross sections  are
smaller than the corresponding numbers for 
$\bar tt\gamma$ selection cuts. Moreover, the cross section for $\bar tt$
production followed by radiative $t$ decay {\it decreases} by a factor~3 to~4 
when
the top quark charge is varied from $q_{top}=2/3$ to $q_{top}=-4/3$.
This is due to interference
effects between the diagrams where the photon is radiated from the
$t$-quark line and the diagrams where the photon is either emitted from
the final state $W$ or the $b$-quark. For
the photon $p_T$ cut used, a significant fraction of the events in the
SM at the Tevatron in the regions defined by the 
$t\to Wb\gamma\to\ell\nu b\gamma$ and $t\to Wb\gamma\to jjb\gamma$ 
selection cuts originates from $\bar tt\gamma$ production. For
$q_{top}=-4/3$, $\bar tt\gamma$ production is in fact the dominant source
in these regions. Nonetheless, the net result is a decrease in the total cross 
section when we switch from $q_{top}=2/3$ to $q_{top}=-4/3$.
Recall that, in contrast,
the cross section increased with the change in top charge
for  $\bar tt\gamma$ selection cuts.  This suggests that the 
{\it ratio} of the cross section associated with radiation in
top production to the radiative decay cross section would be very
sensitive to the top quark charge.  This issue is explored in the next section.

Combining the cross sections for all three phase space regions, we 
expect about 60 $\gamma\ell^\pm\nu b\bar bjj$ events at the Tevatron  for
20~fb$^{-1}$.  The number of events expected at the LHC
is about an order of magnitude larger:  550 $\gamma\ell^\pm\nu b\bar bjj$ 
events for 10~fb$^{-1}$.

\begin{figure}
\phantom{x}
\vskip 14.cm
\includegraphics{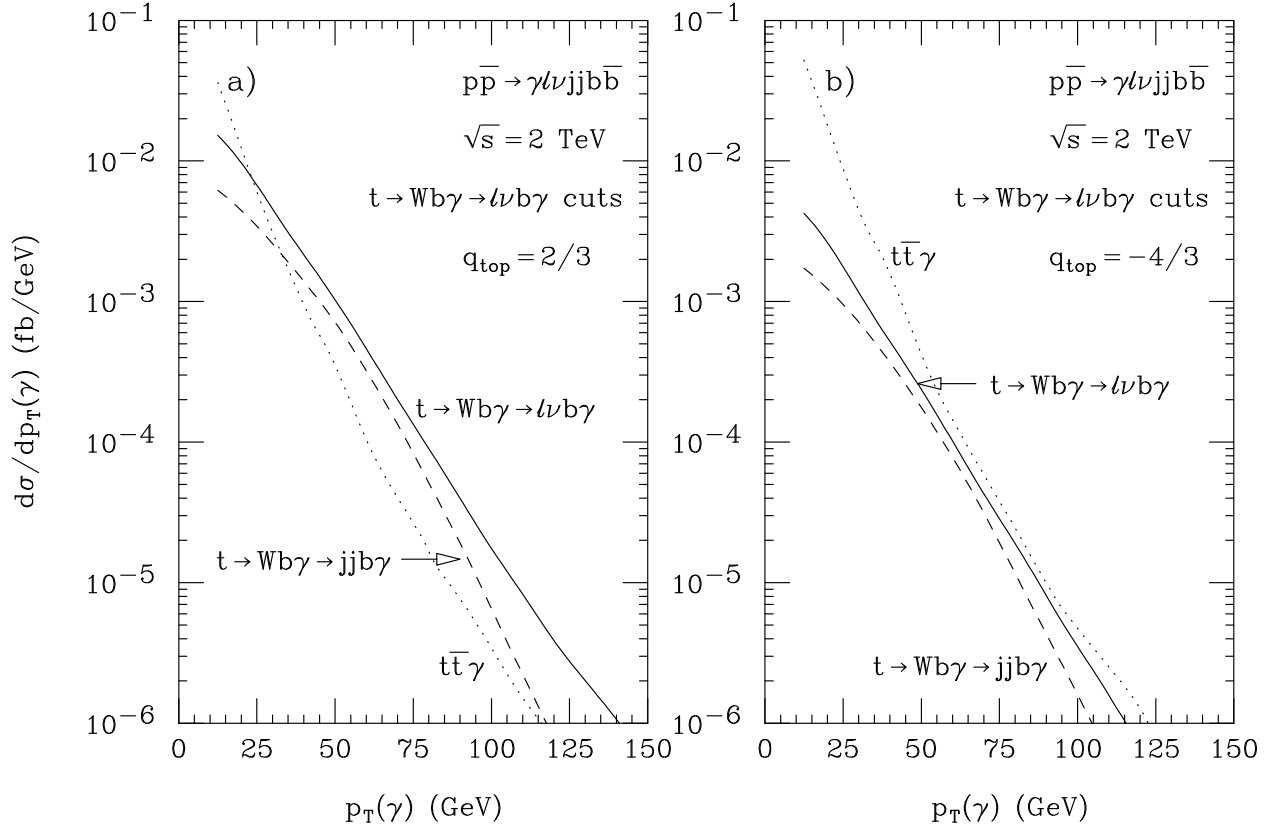}
\caption{The photon transverse momentum distribution for $p\bar p\to
\gamma\ell^\pm\nu b\bar bjj$ at the Tevatron, imposing $t\to 
Wb\gamma\to\ell\nu b\gamma$ selection cuts (see text), for
a) $q_{top}=2/3$ and b) $q_{top}=-4/3$. Shown
are the $\bar tt$, $t\to Wb\gamma\to\ell\nu b\gamma$ (solid line), $\bar
tt$, $t\to Wb\gamma\to jjb\gamma$ (dashed line) and $\bar tt\gamma$
(dotted line) contributions. The additional cuts imposed are described
in Sec.~II.}
\label{fig:one}
\end{figure}

\begin{figure}
\phantom{x}
\vskip 14.cm
\includegraphics{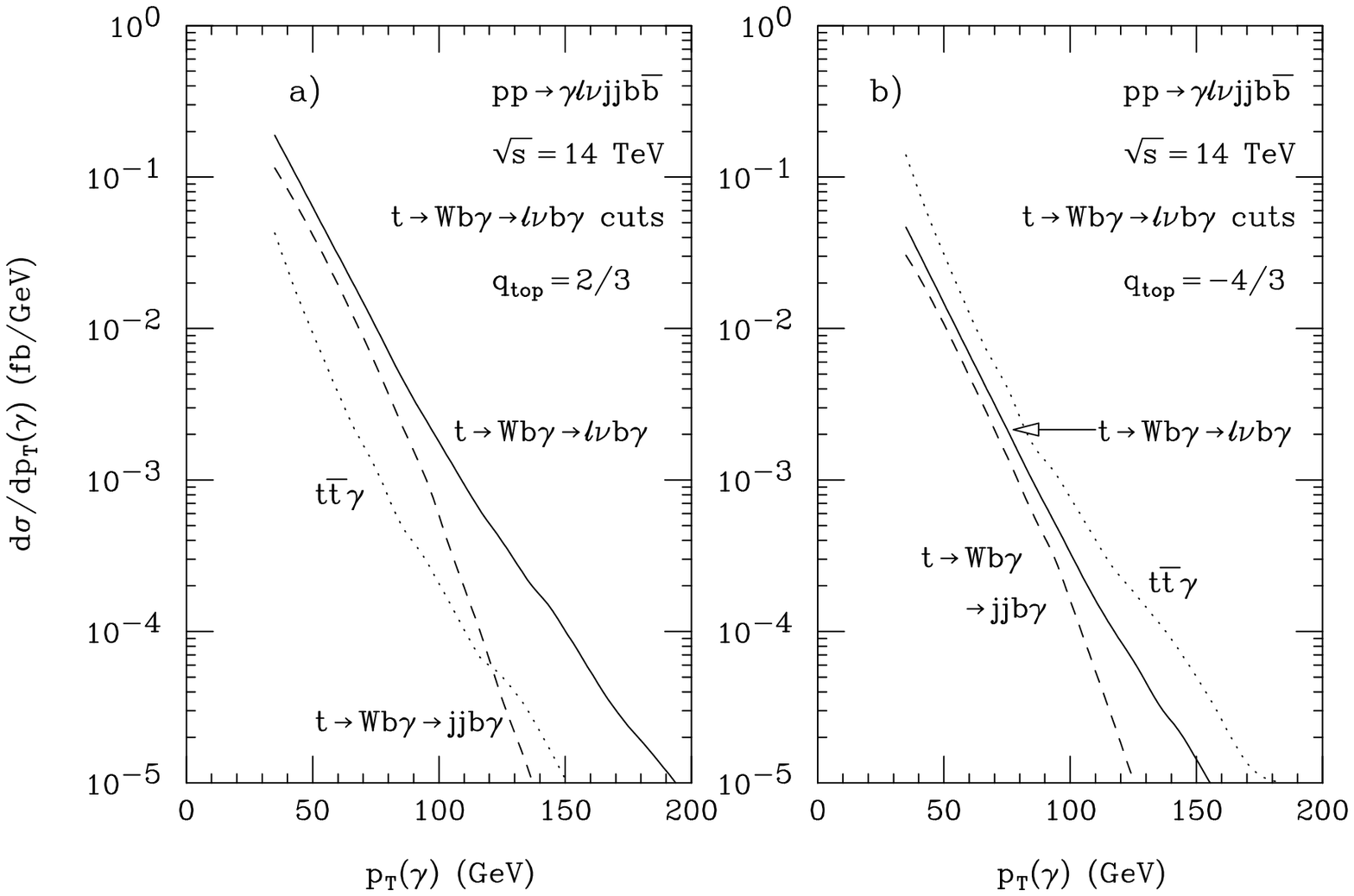}
\caption{The photon transverse momentum distribution for $pp\to
\gamma\ell^\pm\nu b\bar bjj$ at the LHC, imposing $t\to Wb\gamma
\to\ell\nu b\gamma$ selection cuts (see text), for a) 
$q_{top}=2/3$ and b) $q_{top}=-4/3$. Shown
are the $\bar tt$, $t\to Wb\gamma\to\ell\nu b\gamma$ (solid line), $\bar
tt$, $t\to Wb\gamma\to jjb\gamma$ (dashed line) and $\bar tt\gamma$
(dotted line) contributions. The additional cuts imposed are described
in Sec.~II.}
\label{fig:two}
\end{figure}

Returning to the radiative decay sample,
we show in Fig.~\ref{fig:one} the individual contributions to the 
$\gamma\ell^\pm\nu b\bar bjj$ 
differential cross section for $t\to Wb\gamma\to\ell\nu b\gamma$ 
selection cuts at the Tevatron for $q_{top}=2/3$ and 
$q_{top}=-4/3$ as a function of the photon $p_T$. The individual
contributions at the LHC are shown in Fig.~\ref{fig:two}. 
We see that 
the contribution from $\bar tt\gamma$ production in the SM at the
Tevatron exceeds that from $p\bar p\to\bar tt$ with $t\to 
Wb\gamma\to\ell\nu b\gamma$ only for $p_T(\gamma)<25$~GeV 
(Fig.~\ref{fig:one}a). Increasing
the photon $p_T$ cut thus improves the efficiency of the $t\to 
Wb\gamma\to\ell\nu b\gamma$ selection cuts, at the cost of a reduced
event rate. 
Because the cross sections for the radiative top
decay processes decrease whereas the $\bar tt\gamma$ rate increases 
when $q_{top}=-4/3$, 
$\bar tt\gamma$ production is the largest contribution 
to the differential cross section over the entire $p_T(\gamma)$ range 
considered   at both the Tevatron and the LHC (Figs.~\ref{fig:one}b and
\ref{fig:two}b. The 
differential cross section of the $t\to Wb\gamma
\to jjb\gamma$ ``background'' drops much faster with $p_T(\gamma)$ than those
of the two other contributions for large photon transverse momenta. 
When $t\to Wb\gamma\to jjb\gamma$ selection
cuts are imposed,
qualitatively similar results are obtained, with the roles
of the $t\to Wb\gamma\to\ell\nu b\gamma$ and $t\to Wb\gamma\to jjb\gamma$
curves exchanged and somewhat reduced contributions from $\bar tt\gamma$
production.

\begin{figure}
\phantom{x}
\vskip 14.cm
\includegraphics{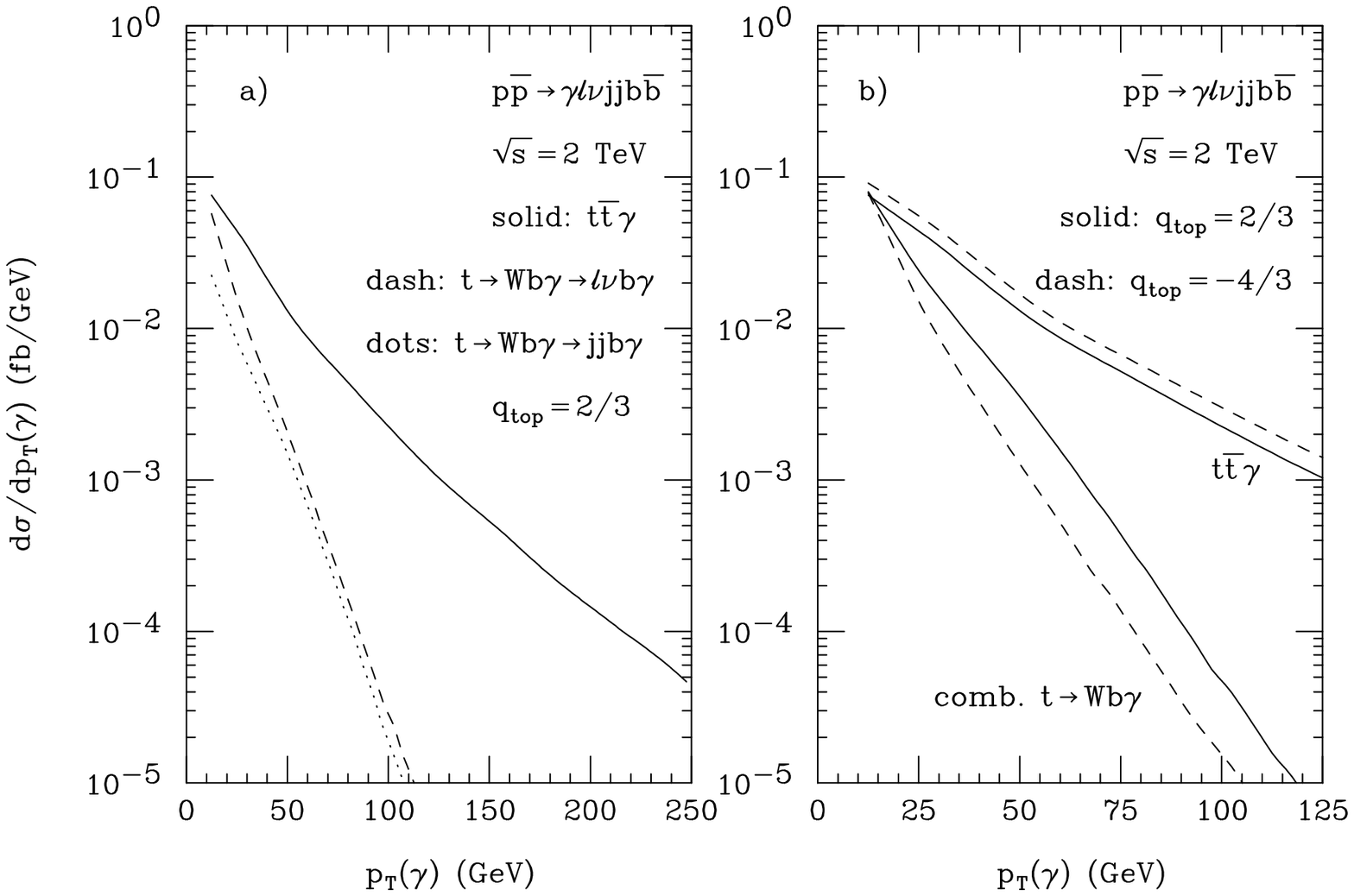}
\caption{The photon transverse momentum distribution for $p\bar p\to
\gamma\ell^\pm\nu b\bar bjj$ at the Tevatron. Part a) displays the
individual contributions for $q_{top}=2/3$, imposing $\bar tt\gamma$ 
(solid line), 
$t\to Wb\gamma\to\ell\nu b\gamma$ (dashed line) and $t\to Wb\gamma\to
jjb\gamma$ selection cuts (dotted line). In part~b), the photon $p_T$ 
distributions for the phase space region defined by the $\bar tt\gamma$ 
selection cuts and the two radiative top decay regions combined are shown 
for $q_{top}=2/3$ and $q_{top}=-4/3$. The cuts imposed are described in 
the text.}
\label{fig:three}
\end{figure}

\begin{figure}
\phantom{x}
\vskip 14.cm
\includegraphics{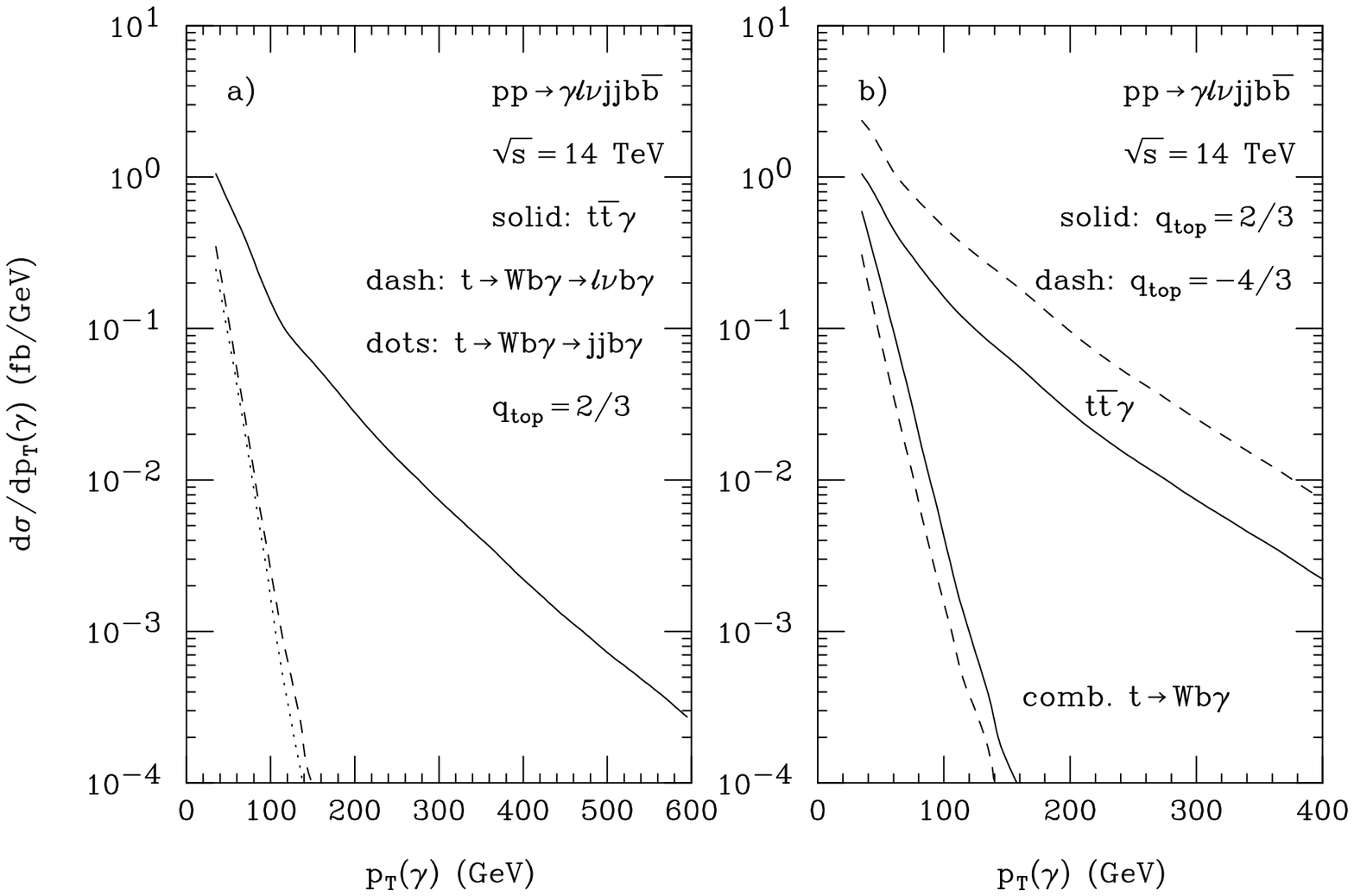}
\caption{The photon transverse momentum distribution for $pp\to
\gamma\ell^\pm\nu b\bar bjj$ at the LHC. Part~a) displays the
individual contributions for $q_{top}=2/3$, imposing $\bar tt\gamma$ 
(solid line), 
$t\to Wb\gamma\to\ell\nu b\gamma$ (dashed line) and $t\to Wb\gamma\to
jjb\gamma$ selection cuts (dotted line). In part b), the photon $p_T$ 
distributions for the phase space region defined by the $\bar tt\gamma$ 
selection cuts and the two radiative top decay regions combined are shown 
for $q_{top}=2/3$ and $q_{top}=-4/3$. The cuts imposed are described in 
the text.}
\label{fig:four}
\end{figure}

The preceding results are summarized in Figs.~\ref{fig:three}
and~\ref{fig:four}.  Fig.~\ref{fig:three}a shows the total 
SM ($q_{top}=+2/3$) cross section for $p\bar p\to \gamma\ell^\pm\nu
b\bar bjj$ at the  
Tevatron, with all contributions from radiation in top production and
radiative decays combined.  The three curves correspond to the three
sets of selection cuts.  Fig.~\ref{fig:four}a shows the corresponding 
results for the LHC (with $pp$ initial state).  We see in these figures, as
we have already seen
in Table~\ref{tab:one}, that the $\bar tt\gamma$ selection cuts (solid curves) 
result in higher cross sections at both colliders than the cuts
for radiative decays $t\to Wb\gamma$.  In Figs.~\ref{fig:three}a
and~\ref{fig:four}a we also see that the photon distribution 
imposing $t\to Wb\gamma\to\ell\nu b\gamma$ and
$t\to Wb\gamma\to jjb\gamma$ selection cuts  drops much faster than
the $p_T(\gamma)$ distribution in the phase space region defined by the
$\bar tt\gamma$ selection cuts.  This is a consequence of the 
phase space restriction on the photon energy in radiative top
quark decays.


In Figs.~\ref{fig:three}b and~\ref{fig:four}b  we compare cross sections
for $q_{top}=+2/3$ (solid curves) and $q_{top}=-4/3$ (dashed curves) for the 
Tevatron and LHC, respectively.
In each figure the upper pair of curves  corresponds to $\bar tt\gamma$ 
selection cuts 
and the lower pair includes events from both radiative decay regions
combined.  The figures show that at both machines, the effect of 
increasing $|q_{top}|$ is to increase the cross section in the $\bar tt\gamma$
cuts region and to decrease it in the radiative decay regions, for all
values of photon $p_T$.  At the Tevatron, however, 
since the contributions from radiative top quark decays
are small if $\bar tt\gamma$ selection cuts are imposed, changing the 
top quark charge has little influence on the shape of the photon $p_T$
distribution in this region (upper curves in Fig.~\ref{fig:three}b). 
On the other hand, at small values of
$p_T(\gamma)$, $\bar tt\gamma$ production may contribute significantly to
the cross section in the regions defined by the $t\to Wb\gamma\to\ell\nu
b\gamma$ and $t\to Wb\gamma\to jjb\gamma$ selection cuts (see
Table~\ref{tab:one}). The shape of
the photon transverse momentum distribution in these regions thus may
be sensitive to $q_{top}$ (lower curves in Fig.~\ref{fig:three}b). 
The curves at the LHC (Fig.~\ref{fig:four}b) also exhibit some shape 
dependence, albeit less pronounced than at the Tevatron.

\begin{figure}
\phantom{x}
\vskip 14.cm
\includegraphics{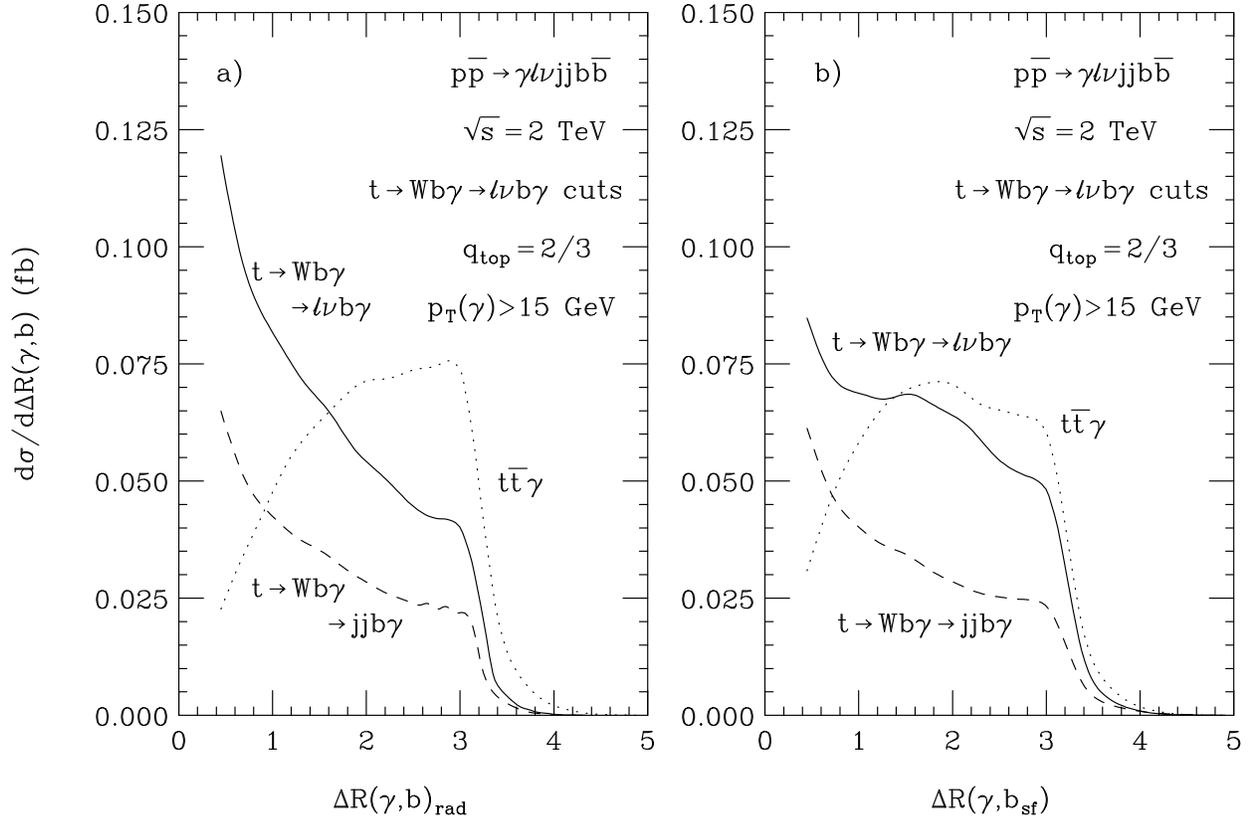} 
\caption{Comparison of the SM $\Delta R(\gamma,b)_{rad}$ and $\Delta
R(\gamma,b_{sf})$ distributions for $p\bar p\to\gamma\ell\nu jjb\bar b$
at the Tevatron, imposing $t\to Wb\gamma\to\ell\nu b\gamma$ selection
cuts (see text). Shown are the contributions from $\bar 
tt$ production with $t\to Wb\gamma\to\ell\nu b\gamma$ (solid lines) and 
$t\to Wb\gamma\to jjb\gamma$ (dashed lines), and $\bar
tt\gamma$ (dotted lines) production. The cuts described in Sec.~II are 
imposed, except for the minimum photon transverse momentum which has
been increased to 15~GeV.}
\label{fig:five}
\end{figure}

We can look for sensitivity to $q_{top}$ in other distributions as well.
In $t\to Wb\gamma$, emission of the photon from the $b$-quark line leads
to a collinear enhancement at small values of the separation between the
photon and the $b$-quark, $\Delta R(\gamma,b)_{rad}$, which depends on
$q_b=q_{top}-q_W$. Therefore, in radiative top decays, the shape of the 
$\Delta R(\gamma,b)_{rad}$
distribution is expected to be sensitive to the charge of the top
quark. In contrast, the photon and the $b$-quark originating from the 
$t\to Wb$ decay in $\bar tt\to\gamma W^+bW^-\bar b$ events are
essentially uncorrelated. In $\bar tt\gamma$
events, the photon and {\sl both} $b$-quarks are largely
uncorrelated. Unfortunately, on an event by event basis, it is difficult
to determine whether the $b$ or the $\bar b$ quark originates from the 
radiative top decay. However, due to the collinear enhancement for small
values of $\Delta R$ in radiative top decays, 
\begin{equation}
\Delta R(\gamma,b)_{min}=\min[\Delta R(\gamma,b),\Delta R(\gamma,\bar
b)]
\end{equation}
should coincide with the photon -- $b$-quark separation in radiative top
decays for small values
of $\Delta R$. Alternatively, the $\Delta R(\gamma,b_{sf})$ distribution, where
$b_{sf}$ is the $b$ or $\bar b$-quark with the smaller transverse momentum 
\begin{equation}
p_T(b_{sf})=\min[p_T(b),p_T(\bar b)]
\end{equation}
may be considered. Since a $b$-quark which radiates a photon
looses energy, the $b_{sf}$ in $\bar tt\to\gamma W^+bW^-\bar b$ 
frequently originates from the radiative top decay. 
It turns out that the $\Delta R(\gamma,b_{sf})$
distribution more closely resembles the true $\Delta R(\gamma,b)_{rad}$
distribution. Figure~\ref{fig:five} shows a comparison of the $\Delta
R(\gamma,b)_{rad}$ and $\Delta R(\gamma,b_{sf})$
distributions at the Tevatron in the SM case, imposing $t\to Wb\gamma 
\to\ell\nu b\gamma$ selection cuts. In each
case, the individual distributions of the three contributing
subprocesses are shown. Qualitatively
similar results are obtained for the LHC case, and for $t\to Wb\gamma
\to jjb\gamma$ selection cuts. Due to the collinear enhancement at small
opening angles between the $b$-quark and the photon, the 
$\Delta R(\gamma,b)_{rad}$ distribution rises very quickly with 
decreasing values of $\Delta R(\gamma,b)_{rad}$ for the radiative top 
decay processes. For $\bar tt\gamma$ production no collinear enhancement
occurs for either the $\Delta R(\gamma,b)$ or the $\Delta R(\gamma,\bar b)$
distribution. In Fig.~\ref{fig:five}a, we have chosen the $\Delta 
R(\gamma,\bar b)$ distribution 
for $\bar tt\gamma$ production (dotted line). Very similar results
are obtained if the $\Delta R(\gamma,b)$ distribution is chosen
instead. Except for the steepness of the distribution at small 
$\Delta R(\gamma,b_{sf})$ for $t\to Wb\gamma\to\ell\nu b\gamma$, the shapes
of the $\Delta R(\gamma,b)_{rad}$ and $\Delta R(\gamma,b_{sf})$ 
distributions are found to agree reasonably well. 

In Fig.~\ref{fig:five} we have increased the 
minimum photon transverse momentum requirement of Eq.~(\ref{eq:top4}) to 
$p_T(\gamma)>15$~GeV. 
As we have mentioned before, $\bar tt\gamma$ production contributes
significantly to the cross section at the Tevatron in the region 
defined by the $t\to Wb\gamma \to\ell\nu b\gamma$ selection cuts for 
the cuts specified in Sec.~II. The $\bar tt\gamma$ contribution can be
reduced by increasing the $p_T$ cut on the photon (see
Fig.~\ref{fig:one}), at the cost of a decreased number of
events. Increasing the  photon transverse momentum cut to
$p_T(\gamma)>15$~GeV reduces the $\bar tt\gamma$ cross section for $t\to
Wb\gamma \to\ell\nu b\gamma$ selection cuts by about a factor~2, whereas
the contributions involving radiative top decays decrease by only 30\%. 

\begin{figure}
\phantom{x}
\vskip 14.cm
\includegraphics{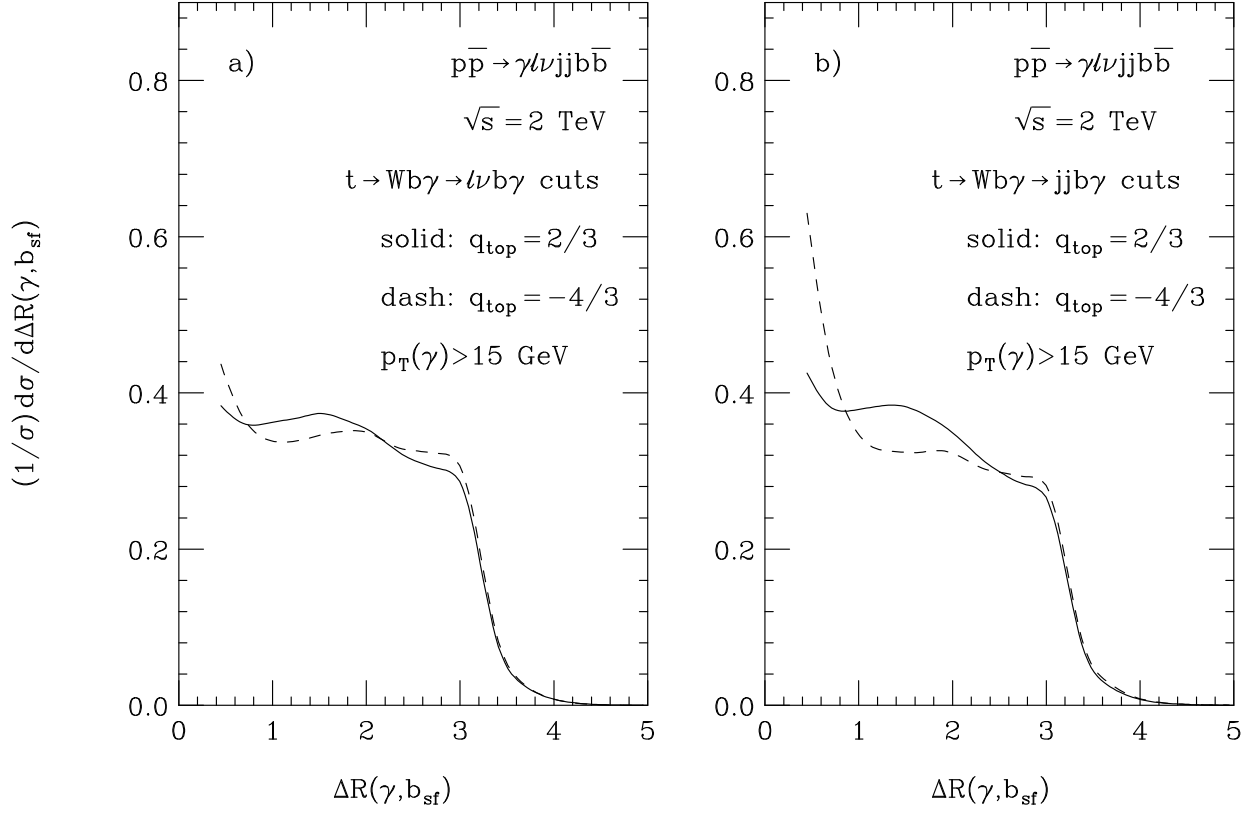} 
\caption{The normalized $(1/\sigma)\,(\Delta R(\gamma,b_{sf}))$ 
distribution for $p\bar p\to\gamma\ell\nu jjb\bar b$
at the Tevatron, imposing a) $t\to Wb\gamma\to\ell\nu b\gamma$ selection
cuts and b) $t\to Wb\gamma\to jjb\gamma$ selection
cuts (see text). Shown are the results for a SM top quark
charge assignment (solid lines) and for $q_{top}=-4/3$ (dashed
lines). The cuts described in Sec.~II are imposed, except for the
minimum photon transverse momentum which has been increased to 15~GeV.}
\label{fig:six}
\end{figure}

\begin{figure}
\phantom{x}
\vskip 14.cm
\includegraphics{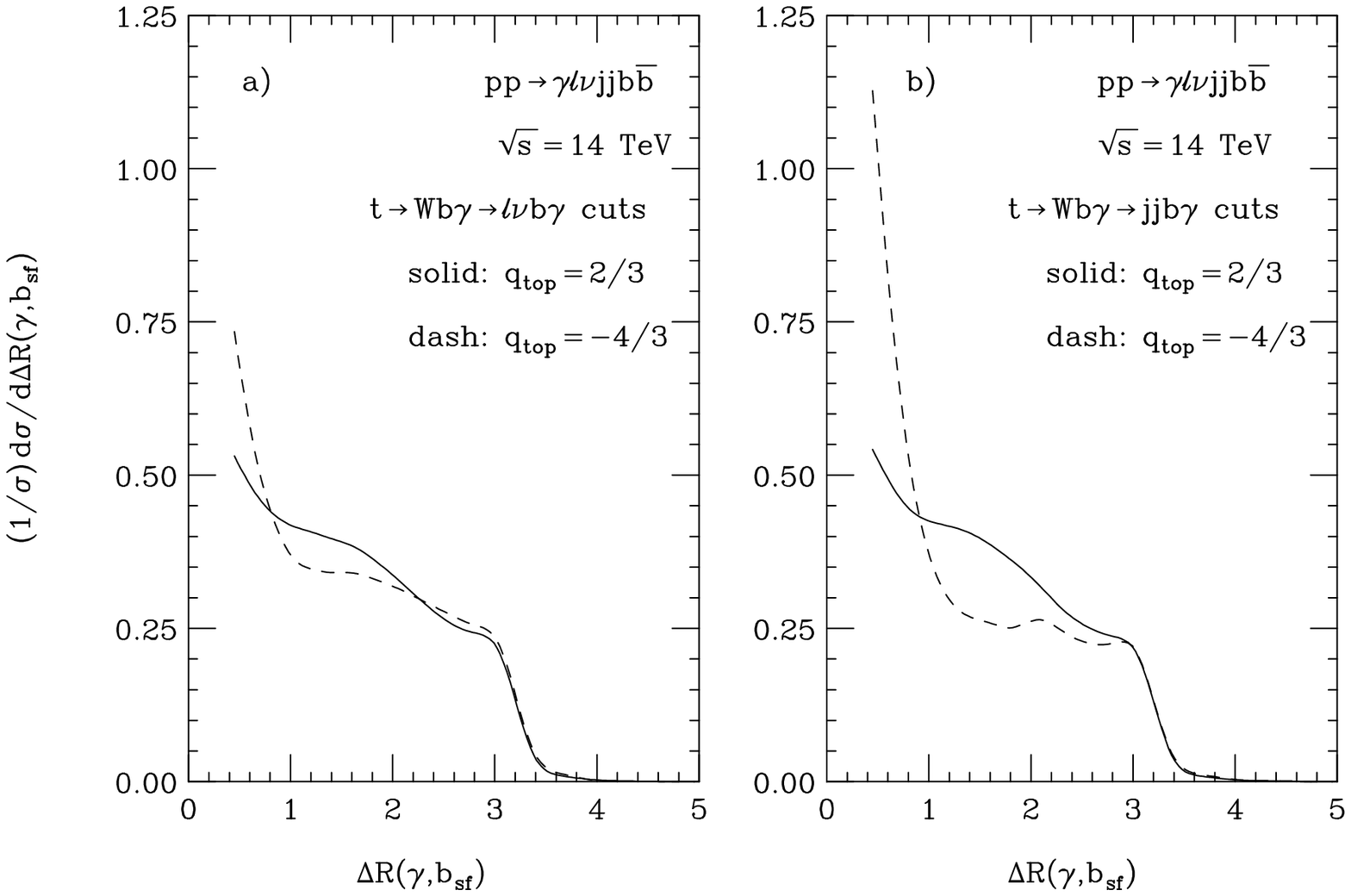} 
\caption{The normalized $(1/\sigma)\,(\Delta R(\gamma,b_{sf}))$ 
distribution for $pp\to\gamma\ell\nu jjb\bar b$
at the LHC, imposing a) $t\to Wb\gamma\to\ell\nu b\gamma$ selection
cuts and b) $t\to Wb\gamma\to jjb\gamma$ selection
cuts (see text). Shown are the results for a SM top quark
charge assignment (solid lines) and for $q_{top}=-4/3$ (dashed lines). 
The additional cuts imposed are described in Sec.~II.}
\label{fig:seven}
\end{figure}

We compare the $\Delta R$ distributions for the two top quark charges
in Figures~\ref{fig:six} and~\ref{fig:seven}.
We display the normalized
distribution $(1/\sigma)\,(d\sigma/d\Delta R(\gamma,b_{sf}))$ for
$\bar tt\to\gamma\ell\nu jj\bar bb$ with
$q_{top}=2/3$ and $q_{top}=-4/3$ at the Tevatron and LHC,
respectively. Normalizing the distributions to unit area emphasizes the 
dependence of the shape of the $\Delta R(\gamma,b_{sf})$
distribution to the top quark charge. For a charge $-4/3$ quark,
the $\Delta R(\gamma,b_{sf})$ distribution rises relatively faster in
the region $\Delta R(\gamma,b_{sf})\leq 1$, but is flatter for 
$1<\Delta R(\gamma,b_{sf})<3$, than if $q_{top}=2/3$. At small values 
of $\Delta R(\gamma,b_{sf})$, emission from the
$b$-quark line dominates. Since the {\sl magnitude} of $q_b$ is the same
for $q_{top}=2/3$ and $q_{top}=-4/3$, the differential cross section is
similar in both cases. However, the total cross section for
$q_{top}=-4/3$ within cuts is significantly smaller than for a SM top
quark charge assignment (see Table~\ref{tab:one}), due to destructive 
interference between the contributing Feynman diagrams. This leads to a
larger normalized differential cross section for $q_{top}=-4/3$ at small
$\Delta R(\gamma,b_{sf})$. In the region $1<\Delta R(\gamma,b_{sf})<3$,
the destructive interference is responsible
for the flatness of the $(1/\sigma)\,(d\sigma/d\Delta R(\gamma,b_{sf}))$
distribution in the $q_{top}=-4/3$ case. For $\Delta
R(\gamma,b_{sf})>3$, the $b$-quark with the
smaller $p_T$ most of the time coincides with the bottom quark from the
(regular) $t\to Wb$ decay and, thus, carries little information on
$q_{top}$. The shape of the $\Delta R(\gamma,b_{sf})$ distribution in
the phase space region defined by the $t\to Wb\gamma\to jjb\gamma$
selection cuts 
(Figs.~\ref{fig:six}b and~\ref{fig:seven}b) is found to be somewhat more
sensitive to $q_{top}$ than in the case when $t\to Wb\gamma\to\ell\nu b\gamma$
selection cuts are imposed (Figs.~\ref{fig:six}a
and~\ref{fig:seven}a). At the Tevatron (Fig.~\ref{fig:six}), we have 
again imposed a
$p_T(\gamma)>15$~GeV cut. The shape of the $\Delta R(\gamma,b_{sf})$
distribution for $\bar tt\to\gamma W^+bW^-\bar b$ at the LHC is in
general more sensitive to the top quark charge than at the Tevatron. 

\section{Quantitative Estimates}

In this section we determine quantitative bounds on 
\begin{equation}
\Delta q_{top}=q_{top}-{2\over 3}
\end{equation}
which one expects to achieve with 20~fb$^{-1}$ of data
at the Tevatron in Run~II, and 10~fb$^{-1}$ at the LHC, analyzing the
photon $p_T$ and the $\Delta R(\gamma,b_{sf})$ distributions. We also
briefly consider the ratio of the cross section
associated with radiation in top production to the radiative decay cross
section which may be useful for small data sets.

In our analysis of differential cross sections, we calculate
95\% confidence level (C.L.) limits performing a $\chi^2$ test. The
statistical significance is calculated by splitting the selected
distribution into a number of bins, each with typically more than five
events. In each bin the Poisson statistics are approximated by a Gaussian
distribution. We use the photon transverse momentum distribution in the phase
space region defined by the $\bar tt\gamma$ selection cuts 
of~(\ref{eq:top51}), and the $p_T(\gamma)$ and $\Delta
R(\gamma,b_{sf})$ distributions of the combined radiative top decay 
regions, defined by requiring that events satisfy~(\ref{eq:top6}) 
or~(\ref{eq:top7}) but not~(\ref{eq:top51}). We impose
the cuts described in Sec.~II, except for the $\Delta
R(\gamma,b_{sf})$ distribution, where we replace the $p_T$ cut on the
photon of Eq.~(\ref{eq:top4}) with $p_T(\gamma)>15$~GeV.
We combine channels with electrons and muons (both charge
combinations) in the final state. We also make use of the approximate 
symmetry of the cross
sections under the transformation~(\ref{eq:trans}) and define the $b$-quark
charge to be $q_b=q_{top}-1$. Confidence levels calculated for
$q_{top}=+4/3$ will thus automatically apply for a charge $-4/3$ quark
which decays into a $b$-quark and a $W^-$. 

Our calculation does not include QCD corrections which are expected to
modify the cross section by $20-40\%$. 
In order to derive realistic limits, we therefore allow for a normalization
uncertainty of $\Delta{\cal N}=30\%$ of the SM cross section. The 
expression for $\chi^2$
which is then used to compute confidence levels is given by~\cite{babe}
\begin{equation}
\chi^2=\sum_{i=1}^{n_D}\,{(N_i-fN^0_i)^2\over fN_i^0}+(n_D-1)\, ,
\end{equation}
where $n_D$ is the number of bins, $N_i$ is the number of events for a
given $\Delta q_{top}$, and $N_i^0$ is the number of events in the SM in
the $i$th bin. $f$ reflects the uncertainty in the normalization of the
SM cross section within the allowed range, and is determined by
minimizing $\chi^2$:
\begin{equation}
f=\cases{(1+\Delta{\cal N})^{-1} & for $\bar f < (1+\Delta{\cal N})
^{-1}$ \cr
\bar f & for $ (1+\Delta{\cal N})^{-1}<\bar f<1+\Delta{\cal N} $ \cr
1+\Delta{\cal N} & for $\bar f > 1+\Delta{\cal N}$} 
\end{equation}
with
\begin{equation}
\bar f^2=\left\{\sum_{i=1}^{n_D}N_i^0\right\}^{-1}\,\sum_{i=1}^{n_D}{N_i^2
\over N_i^0}~.
\end{equation}

In Table~\ref{tab:two} we display the resulting 95\% confidence level 
limits for $\Delta q_{top}$ expected from the Tevatron Run~II for an 
integrated luminosity of 20~fb$^{-1}$, and from the LHC for
10~fb$^{-1}$. The last row shows the combined limits from the three
individual distributions studied here.
Possible backgrounds are assumed to be subtracted.
At both the Tevatron and the LHC, the $p_T(\gamma)$ and $\Delta
R(\gamma,b_{sf})$ distributions in the combined radiative top decay
regions yield rather similar limits for $\Delta q_{top}$. As mentioned
before, $\bar tt\gamma$ production at the Tevatron is completely
dominated by $q\bar q$ annihilation and photon radiation from the
initial state quarks limits the sensitivity of the cross section to
$q_{top}$. 
Therefore, the bounds which can be achieved at the Tevatron 
from analyzing the photon $p_T$ distribution in the phase space region 
defined by the $\bar tt\gamma$ selection cuts are significantly weaker than
those which can be obtained from either the $p_T(\gamma)$ or the $\Delta
R(\gamma,b_{sf})$ distribution in the combined radiative top decay
regions.\footnote{In practice, hadronization effects we have omitted will come 
into play, and are likely to affect the $\Delta R$ distributions more 
than those for the photon $p_T$.}
 At the LHC, most of the $\bar tt\gamma$ cross section 
originates from gluon
fusion, and the limits from the photon $p_T$ distribution for $\bar
tt\gamma$ selection cuts are similar to those which can be
achieved in the phase space regions where radiative top decays
dominate. 

While a precise measurement of the top quark charge will not be possible
in Run~II of the Tevatron, even with an integrated luminosity of 
20~fb$^{-1}$, we
note that a charge $-4/3$ top quark can be ruled out at $\approx 95\%$
C.L. when the limits from the different distributions and phase space
regions are combined. At the LHC, with 10~fb$^{-1}$, a
measurement of the top quark charge with an accuracy of about 10\%
appears to be feasible. As we pointed out in Sec.~II, the $p_T$ cuts we
use are suitable for running at low luminosity at the LHC, i.e. for ${\cal
L}=10^{33}\,{\rm cm^{-2}\,s^{-1}}$. At high luminosity, ${\cal
L}=10^{34}\,{\rm cm^{-2}\,s^{-1}}$, the transverse momentum cuts must be
increased due to pile-up effects caused by the large number of
interactions per beam crossing. The $\gamma\ell^\pm\nu b\bar bjj$ cross
section is quite sensitive to these cuts. 
For example, increasing the $p_T$ cut of
the $b$-quarks to 30~GeV and the photon, jet and missing $p_T$ cuts to
50~GeV reduces the $\bar tt\gamma$ and $\bar tt\to\gamma W^+bW^-\bar b$
cross sections by a factor $\approx 10$ and $\approx 100$,
respectively. It is thus not clear whether running at high luminosity
will improve the measurement of the top quark electric charge at the LHC.

The bounds on $\Delta q_{top}$ are in general 
quite sensitive to $\Delta{\cal N}$. If the normalization uncertainty
can be reduced to $\Delta{\cal N}=10\%$, the limits listed 
in Table~~\ref{tab:two} can be improved by up to a factor~2. Further
improvements may result from using more powerful statistical tools than
the simple $\chi^2$ test we performed. 

For integrated luminosities smaller than 20~fb$^{-1}$ at the Tevatron, 
the limited number of events
makes a detailed analysis of differential cross sections difficult. 
In this case, the ratio of the cross section
associated with radiation in top production to the radiative decay cross
section, ${\cal R}$, may be more useful. Many experimental and theoretical
uncertainties, for example, those associated with the particle 
detection efficiencies, the uncertainty in the integrated
luminosity, or the uncertainties from the choice of scale or of the parton
distribution functions, are expected to cancel, at least partially, in
${\cal R}$. The cross section ratio is quite sensitive to 
$q_{top}$. From the individual cross sections listed in
Table~\ref{tab:one} one finds ${\cal R}=1.73$ for $q_{top}=2/3$, and
${\cal R}=2.88$ for $q_{top}=-4/3$ at the Tevatron. For an integrated
luminosity of 10~fb$^{-1}$, the expected  statistical uncertainty at
the Tevatron is $\Delta {\cal R}_{stat}=0.65$ and a charge $-4/3$ 
top quark can thus be ruled out at $\approx 90\%$~C.L. using the cross
section ratio, if one ignores possible systematic uncertainties. 
At the LHC, the value of ${\cal R}$ is even more sensitive to $q_{top}$
than at the Tevatron, but enough events are expected to allow 
a more detailed analysis.

\section{Summary and Conclusions}

Like most of its fundamental quantum numbers, the electric charge of 
the top quark has not been measured so far. Alternative interpretations
for the particle we believe is the charge $2/3$ isospin partner of the
$b$-quark are thus not ruled out. For example, since the correlations
of the $b$-quarks and the $W$-bosons in $\bar pp\to\bar tt\to W^+W^-\bar
bb$ are not determined, it is conceivable that the ``$t$-quark''
observed at the Tevatron is an exotic quark, $Q_4$, with charge $-4/3$ 
which decays via $Q_4\to W^-b$. This interpretation is consistent with
current precision electroweak data~\cite{ccm}. 
In order to determine the charge of the top quark, one can either
measure the charge of the $b$-jet, or investigate photon radiation in
$\bar tt$ events.  
The latter method actually measures a combination of the EM coupling
strength and 
the charge quantum number. Combining the results of the two methods 
will thus make it possible to determine both quantities.

In this paper we have explored the possibility of measuring $q_{top}$ in 
$p\,p\hskip-7pt\hbox{$^{^{(\!-\!)}}$} \to \gamma\ell^\pm\nu\bar bbjj$, 
assuming that both $b$-quarks are tagged. Our analysis makes use of both
$\bar tt\gamma$ production and radiative top quark decays, 
and is carried out at the tree level. The EM coupling strength is
assumed to take its SM value. Top quark and $W$ boson decays are
treated in the narrow width approximation. Contributions from radiative
$W$ decays can be suppressed by simple phase space cuts (see
Eq.~(\ref{eq:radw})). The remaining
event sample can be separated into a $\bar tt\gamma$ sample, a $\bar
tt$, $t\to Wb\gamma\to\gamma\ell\nu b$ sample, and a $\bar
tt$, $t\to Wb\gamma\to\gamma jjb$ sample by imposing invariant mass and
cluster transverse mass cuts (see Eqs.~(\ref{eq:top51})
--~(\ref{eq:top7})).  

In the phase space region dominated by $\bar tt\gamma$ production, 
the cross section increases if the magnitude of 
$q_{top}$ increases. At the Tevatron, $q\bar q$ annihilation dominates
and photon radiation off the incoming quarks constitutes an irreducible
background which limits the sensitivity to $q_{top}$. In contrast, at
the LHC, gluon fusion dominates, and the $\bar tt\gamma$ cross section
scales approximately with $q^2_{top}$. 

In the phase space regions which enhance $\bar tt$ production with one
of the top quarks decaying radiatively ($t\to Wb\gamma$), interference
effects between the Feynman diagrams where the photon is emitted from
the top quark, the $W$ boson and the $b$-quark line may cause the cross
section to either decrease or increase with $\Delta
q_{top}=q_{top}-2/3$. For example, for $q_{top}=-4/3$, the
$\gamma\ell^\pm\nu\bar bbjj$ cross section via radiative top quark
decays is reduced by a factor of $3$ to $4$. 

The changes in the individual cross sections are reflected in the photon
transverse momentum distribution in the various phase space regions. 
We also found that the 
$\Delta R(\gamma,b_{sf})$ distribution, where $b_{sf}$ is the $b$ or
$\bar b$-quark with the smaller $p_T$, is sensitive to the charge of the
top quark in the phase space regions dominated by $\bar tt$ production 
where one of the top quarks decays radiatively.

In order to determine how well one can hope to measure the top quark
charge using photon radiation in top quark events in future Tevatron 
and LHC experiments, we have performed $\chi^2$ tests of the 
$p_T(\gamma)$ and the $\Delta R(\gamma,b_{sf})$ distributions in the
various phase space regions of interest. At the Tevatron, with an 
integrated luminosity of 20~fb$^{-1}$, one will be able to exclude at 
$\approx 95\%$~C.L. the possibility that an exotic
quark $Q_4$ with charge $-4/3$ and not the SM top quark has been found
in Run~I. For smaller integrated luminosities, the number of events
expected is very small and it will be difficult to perform a 
quantitative analysis using differential cross sections. In this case, 
the ratio of the cross section associated with radiation in top
production to the radiative decay cross section may be a useful tool.
At the LHC with 10~fb$^{-1}$ obtained at ${\cal
L}=10^{33}\,{\rm cm^{-2}\,s^{-1}}$,
it should be possible to measure the electric charge of the top quark
with an accuracy of about 10\%. For comparison, at a linear collider with
$\sqrt{s}=500$~GeV and $\int\!{\cal L}dt=200$~fb$^{-1}$, one expects
that $q_{top}$ can also be measured with a precision of about
10\%~\cite{resource}. Finally, at a $\gamma\gamma$ collider it 
is conceivable that the top quark charge can be determined with an
accuracy of better than 1\%, if the $\gamma\gamma\to\bar tt$ cross section
can be measured with a precision of 2\%~\cite{boos}. 

The measurement of the electric charge of the top quark at the 
Tevatron using photon radiation in top quark events is severely limited 
by statistics. Even with an integrated luminosity of 20~fb$^{-1}$,
$q_{top}$ can only be determined with a precision of
$30-100\%$. However, as we pointed out in Sec.~I, information on the 
electric charge of the top quark may also be obtained from a 
measurement of the charge of the $b$-jets, and the charge sign of the 
leptons in semileptonically tagged $b$-quarks. This may significantly 
improve the precision which can be obtained. Detailed simulations will 
be necessary in order to determine how well
the top quark charge can be measured using these methods. 

\acknowledgements
We would like to thank C.~Ferreti, H.~Frisch, B.~Knuteson, 
T.~LeCompte, J.~Parsons, Z.~Sullivan and J.~Womersley for stimulating 
discussions. One of us (U.B.) is grateful to the Fermilab Theory Group,
where part of this work was carried out, for its generous hospitality.
This work has been supported in part by DOE
grant~DE-FG02-91ER40685, NSF grants~PHY-9600155 and PHY-9970703,
and a NSF Graduate Student Fellowship. 

%
%

%
\newpage
%
\widetext
\begin{table}
\caption{Integrated cross sections for $\gamma\ell^\pm\nu\bar bbjj$
production at the Tevatron ($p\bar p$ collisions at $\sqrt{s}=2$~TeV)
and the LHC ($pp$ collisions at $\sqrt{s}=14$~TeV) for $q_{top}=2/3$ and
$q_{top}=-4/3$ imposing a)~$\bar tt\gamma$ selection cuts,
b) $t\to Wb\gamma\to\ell\nu b\gamma$ selection cuts,
and c) $t\to Wb\gamma\to jjb\gamma$ selection cuts.
In addition, the cuts described in Sec.~II are
imposed. In each case, the contributions from $\bar tt\gamma$
production, $\bar tt$ production with $t\to Wb\gamma\to\ell\nu b\gamma$, and
$\bar tt$ production with $t\to Wb\gamma\to jjb\gamma$ are shown. } 
\label{tab:one}
\vskip 3.mm
\begin{tabular}{ccccc}
\multicolumn{5}{c}{a) $\bar tt\gamma$ selection cuts}\\ \tableline
 & $\bar tt\gamma$ & $t\to Wb\gamma\to\ell\nu b\gamma$ & $t\to
Wb\gamma\to jjb\gamma$ & \\
 & contribution & contribution & contribution & total \\
\tableline
Tevatron, $q_{top}=+{2\over 3}$ & 1.70~fb & 0.12~fb & 0.10~fb & 1.92~fb\\
Tevatron, $q_{top}=-{4\over 3}$ & 2.37~fb & 0.03~fb & 0.02~fb & 2.42~fb\\
\tableline
LHC, $q_{top}=+{2\over 3}$ & 33.3~fb & 5.78~fb & 4.86~fb & 44.0~fb \\
LHC, $q_{top}=-{4\over 3}$ & 111.4~fb & 1.21~fb & 0.97~fb & 113.6~fb \\ 
\tableline
\multicolumn{5}{c}{b) $t\to Wb\gamma\to\ell\nu b\gamma$ selection cuts}\\ 
\tableline
Tevatron, $q_{top}=+{2\over 3}$ & 0.36~fb & 0.26~fb & 0.13~fb & 0.75~fb \\
Tevatron, $q_{top}=-{4\over 3}$ & 0.52~fb & 0.07~fb & 0.03~fb & 0.62~fb \\
\tableline
LHC, $q_{top}=+{2\over 3}$ & 0.68~fb & 3.67~fb & 2.26~fb & 6.61~fb \\
LHC, $q_{top}=-{4\over 3}$ & 2.25~fb & 0.87~fb & 0.59~fb & 3.71~fb \\ 
\tableline
\multicolumn{5}{c}{c) $t\to Wb\gamma\to jjb\gamma$ selection cuts}\\ 
\tableline
Tevatron, $q_{top}=+{2\over 3}$ & 0.11~fb & 0.06~fb & 0.19~fb & 0.36~fb \\
Tevatron, $q_{top}=-{4\over 3}$ & 0.16~fb & 0.01~fb & 0.05~fb & 0.22~fb \\
\tableline
LHC, $q_{top}=+{2\over 3}$ & 0.15~fb & 1.07~fb & 3.63~fb & 4.85~fb \\
LHC, $q_{top}=-{4\over 3}$ & 0.49~fb & 0.26~fb & 0.85~fb & 1.60~fb\\ 
\end{tabular}
\end{table}

\begin{table}
\caption{Limits achievable at 95\% C.L. for $\Delta q_{top}$ in 
$p\,p\hskip-7pt\hbox{$^{^{(\!-\!)}}$} \to \gamma\ell^\pm\nu\bar bbjj$ at
the Tevatron and the LHC. Bounds are shown for
the photon transverse momentum distribution in the phase
space region defined by the $\bar tt\gamma$ selection cuts, and the 
$p_T(\gamma)$ and $\Delta R(\gamma,b_{sf})$ distributions of the 
combined radiative top decay regions. The last
row displays the combined limits from the three distributions. The $b$-quark
charge is given by $q_b=q_{top}-1$. The cuts imposed are described in
Sec.~II. For the $\Delta R(\gamma,b_{sf})$ distribution, the
$p_T(\gamma)$ cut of Eq.~(\ref{eq:top4}) has been replaced by
$p_T(\gamma)>15$~GeV. }
\label{tab:two}
\vskip 3.mm
\begin{tabular}{ccc}
distribution & Tevatron, $\int\!{\cal L}dt=20~{\rm fb}^{-1}$ &
LHC, $\int\!{\cal L}dt=10~{\rm fb}^{-1}$ \\ \tableline
$p_T(\gamma)$, $\bar tt\gamma$ region & $-1.16 \leq \Delta q_{top}\leq 
1.11$ & $-0.21 \leq \Delta q_{top}\leq 0.24$ \\
$p_T(\gamma)$, comb. $t\to Wb\gamma$ regions &
$-0.32 \leq \Delta q_{top}\leq 1.20$ & $-0.16 \leq \Delta q_{top}\leq
0.18$ \\
$\Delta R(\gamma,b_{sf})$, comb. $t\to Wb\gamma$ regions & $-0.29 \leq \Delta 
q_{top}\leq 1.10$ & $-0.15 \leq \Delta
q_{top}\leq 0.18$ \\ \tableline
combined & $-0.21 \leq \Delta q_{top}\leq 0.65$ & $-0.067 \leq \Delta
q_{top}\leq 0.070$ \\
\end{tabular}
\end{table}

\newpage
%
%
%
\newpage
\newpage
\newpage
\newpage
\newpage
\newpage
\newpage
\newpage
\end{document}